\documentclass[aps,pra,twocolumn,superscriptaddress]{revtex4}

\usepackage{amsmath}
\usepackage{amsfonts}
\usepackage{amssymb}
\usepackage{graphicx}

\providecommand{\U}[1]{\protect\rule{.1in}{.1in}}

\begin{document}

\title{Purity and Gaussianity bounded uncertainty relation}
\author{A.~Mandilara}
\affiliation{Quantum Information and Communication, \'{E}cole Polytechnique, CP~165/59,
Universit\'{e} Libre de Bruxelles, 1050 Brussels, Belgium} 
\affiliation{Universit\'{e} Paris Diderot, Sorbonne Paris Cit\'{e}, Laboratoire
Mat\'{e}riaux et Ph\'{e}nom\`{e}nes Quantiques, CNRS-UMR 7162, Case courrier 7021, 75205
Paris Cedex 13, France}
\author{ E.~Karpov}
\affiliation{Quantum Information and Communication, \'{E}cole Polytechnique, CP~165/59,
Universit\'{e} Libre de Bruxelles, 1050 Brussels, Belgium}
\author{ N.~J.~Cerf}
\affiliation{Quantum Information and Communication, \'{E}cole Polytechnique, CP~165/59,
Universit\'{e} Libre de Bruxelles, 1050 Brussels, Belgium}

\begin{abstract}
Bounded uncertainty relations provide the minimum value of the uncertainty
assuming some additional information on the state. We derive analytically an uncertainty
relation bounded by a pair of constraints, those of purity and Gaussianity.
In a limiting case this uncertainty relation reproduces the purity-bounded
derived by \textit{Man'ko and Dodonov} and the Gaussianity-bounded one 
\textit{[Phys. Rev. A 86, 030102R (2012)].}
\end{abstract}

\maketitle

\affiliation{Quantum Information and Communication, \'{E}cole Polytechnique, CP~165/59,
Universit\'{e} Libre de Bruxelles, 1050 Brussels, Belgium}

\affiliation{Quantum Information and Communication, \'{E}cole Polytechnique, CP~165/59,
Universit\'{e} Libre de Bruxelles, 1050 Brussels, Belgium}

\section{Introduction}

The only states that saturate the Schr\"{o}dinger-Robertson \cite{Robertson}
uncertainty relation for the canonically conjugated coordinates of position
and momenta, are the pure Gaussian states. However, if some additional
information about the state is available then the set of states which
minimize the uncertainty, or else of \textit{minimizing states} (MSs), is
modified while a tighter lower bound on the uncertainty can be derived.

A basic characteristic of a state is its degree of mixedeness. \ The
minimizing set obtained by imposing the constraint of fixed degree of
mixedeness depends on the measure that one chooses to quantify this degree,
i.e. purity, purities of higher order or various entropies (see \cite
{Dodonov} for a review). \ For instance, in the case of the \textit{purity
-bounded uncertainty relation} suggested by Dodonov and Man'ko \cite{Man'ko}
the minimizing set is composed by mixed states, or more precise, mixtures of
number states. On the other hand, in the case of the von-Neumann entropy,
the set of the MSs is composed by the `thermal states' whose temperature is
increasing as the fixed entropy tends to infinity \cite{Bastiaans}. In both
cases, the lower bound on the uncertainty is increasing with the degree of
mixedeness since the mixedeness adds extra `amount' of classical
(statistical) uncertainty.

In a recent work \cite{PRAr} we have suggested an uncertainty relation
bounded by the degree of Gaussianity, a quantity which we introduced in that
same work. There the non-Gaussian MSs are identified for a fixed degree of
Gaussianity and among them \ one finds all the eigenstates of the harmonic
oscillator. Along with the Gaussianity- bounded uncertainty relation, we
have presented a general method for deriving bounded uncertainty relations
that reduces the problem to an eigenvalue problem.

In this work we employ the method exhibited in \cite{PRAr} to derive an
uncertainty relation where the bound depends on two characteristics of the
state, namely its purity and Gaussianity. To our knowledge this is the first
two-dimensional bounded uncertainty relation that has been derived so far.
The uncertainty relation is represented via parametric relations \ which
connect three quantities, namely, the purity, Gaussianity and uncertainty.
This exact expression is difficult to handle analytically and for this
reason we also provide an approximate expression. The derived relation
provides the boundaries for three basic characteristics of a state and can
be used as a tool for visualizing and partitioning the space of non-Gaussian
mixed states.

The structure of the paper is the following. In first step, in Sec.~\ref%
{Section2} we derive the set of MSs of the purity and Gaussianity bounded
uncertainty relation and then in Sec.~\ref{Section3} we provide exact and
approximate expressions for this uncertainty relation. We discuss our and
conclude in Sec.~\ref{Section4}.

\section{Minimizing states\label{Section2}}

Let us start with the position $\hat{x}$ and momentum $\hat{p}$ of\ a
quantum particle in one dimension, which could also be the quadratures of a
single mode of the electromagnetic field, in a state defined by the density
operator $\hat{\rho}$. In its most general form, the Scr\"{o}dinger-Robertson \ (SR) uncertainty relation \cite{Robertson} for the
position and momentum of this particle reads 
\begin{eqnarray}
(\left\langle \hat{x}^{2}\right\rangle -\left\langle \hat{x}\right\rangle
^{2})(\left\langle \hat{p}^{2}\right\rangle -\left\langle \hat{p}\right\rangle ^{2})&  \nonumber \\
-\frac{1}{4}\left( \left\langle \hat{x}\hat{p}+\hat{p}\hat{x}\right\rangle
-2\left\langle \hat{x}\right\rangle \left\langle \hat{p}\right\rangle
\right) ^{2}& \geq \hbar ^{2}/4.  \label{SR}
\end{eqnarray}
The left-hand side is invariant under \textit{linear canonical
transformations }(LCT), i.e. the direct sum of the symplectic
transformations $\mathrm{SL}(2,R)$ and translations $\mathrm{T}(2)$. In
quantum optics language, LCT\ correspond to squeezing, rotations and
displacements, which form the set of Gaussian operations. The invariance of
the uncertainty with respect to LCT becomes directly evident if we express
the left hand side of Eq.(\ref{SR}) in terms of the \ \textit{covariance
matrix} $\mathbf{\gamma }$ \ of the state \ $\hat{\rho}$, defined through
its matrix elements as 
\begin{equation}
\gamma _{ij}\equiv \frac{1}{2}\mathrm{Tr}(\{(\hat{r}_{i}-d_{i}),(\hat{r}_{j}-d_{j})\}\hat{\rho})  \label{Co}
\end{equation}
where $\hat{\mathbf{r}}=(\hat{x},\hat{p})^{T}$, $\mathbf{d}=\mathrm{Tr}(\hat{\mathbf{r}}\hat{\rho})$ is the displacement vector, and $\{\cdot ,\cdot \}$
is the anticommutator. The left hand side of Eq.(\ref{SR}) is simply $\det 
\mathbf{\gamma }$ and therefore is invariant under LCT. For simplicity in
the presentation, we define here the dimensionless quantity 
\begin{equation*}
\alpha \left( \hat{\rho}\right) \equiv \sqrt{\det \mathbf{\gamma }}/(\hbar
/2)
\end{equation*}
which we call \textit{uncertainty}. With this definition the SR uncertainty
relation simply reads\ $\alpha \geq 1$.

The alternative method of derivation of the Scr\"{o}dinger-Robertson
uncertainty relation presented in \cite{PRAr}, exploits the invariance of \
the uncertainty $\alpha $ under LCT. Due to this invariance, it becomes
possible to confine our search of MSs into a specific class of states into
which all states can be reduced under the action of LCT. By constraining \
the MSs to belong to this class, we are led to solve an optimization problem
for $\alpha $ under constraints, which we tackle with Lagrange multipliers'
method. Apart from the constraints of the class, one may impose additional
constraints and thus derive bounded uncertainty relations depending on other
characteristics of the state such as the purity \cite{Man'ko} or the
Gaussianity \cite{PRAr}.

Before we proceed with the derivation of the purity and Gaussianity bounded
uncertainty relation let us first introduce these two quantities. The 
\textit{purity} $\mu $ of a state $\hat{\rho}$ is defined as 
\begin{equation*}
\mu \left( \hat{\rho}\right) \equiv \mathrm{Tr}\left( \hat{\rho}^{2}\right)
\end{equation*}
while the degree of \textit{Gaussianity} is defined \ as \ \cite{PRAr} 
\begin{equation}
g\left( \hat{\rho}\right) \equiv \mathrm{Tr}\left( \hat{\rho}\hat{\rho}
_{G}\right) /\mathrm{Tr}\left( \hat{\rho}_{G}^{2}\right)  \label{g}
\end{equation}
where $\hat{\rho}_{G}$ is a reference Gaussian state uniquely defined by the
mean vector $\mathbf{d}$ and covariance matrix $\mathbf{\gamma }$ of the
state $\hat{\rho}$. The Gaussianity exhibits the following properties (see 
\cite{PRAr} for the proofs):

\textit{(i)} $g$ is invariant under LCT.

\textit{(ii)} $g$ is a bounded quantity, that is, $2/e\leq g\leq 2$, while

\ \ \ \ $g=1$ for Gaussian states (but the converse is not true).

\textit{(iii)} $g$ together with $\alpha $ confines the set of mixed states
with strictly positive Wigner function.

The aim is to find the states that minimize $\alpha $ under the constraints
of fixed $\mu $ and $g$. All three quantities are invariant under LCT and
therefore, as in \cite{PRAr}, without loss of generality we can confine our
search among a specific class of states with covariance matrix proportional
to the unity \ (in Williamson form) and $\mathbf{d=0}$. We should note here
that every state can be reduced in this form under LCT and in this specific
class the reference Gaussian state is just a thermal state \ $\hat{\rho}
_{G}=e^{-\beta \hat{n}}/A$ where $\hat{n}$ is the number operator and $A=\left( \alpha +1\right) /2$ the normalization factor. Our choice to work
within this specific class of states can be translated as constraints on the
state $\hat{\rho}$ 
\begin{align}
\mathrm{Tr}\left( \hat{\rho}\hat{x}\right) & =\mathrm{Tr}\left( \hat{\rho}\hat{p}\right) =0  \label{c1} \\
\mathrm{Tr}\left( \hat{\rho}\left( \hat{x}\hat{p}+\hat{p}\hat{x}\right)
\right) & =\mathrm{Tr}\left( \hat{\rho}\left( \hat{x}^{2}-\hat{p}^{2}\right)
\right) =0.  \label{c2}
\end{align}
In addition we require that the states $\hat{\rho}$ which minimize the
uncertainty \ 
\begin{equation}
\alpha =\mathrm{Tr}\left( \hat{\rho}\ \left( 2\hat{n}+1\right) \right)
\label{q3}
\end{equation}
are of fixed purity and Gaussianity degree,

\begin{align}
\mu & =\mathrm{Tr}\left( \hat{\rho}^{2}\right)  \label{c3} \\
g & =\frac{1}{N}\mathrm{Tr}\left( \hat{\rho}e^{-\beta\hat{n}}\right)
\label{c4}
\end{align}
where $\mathrm{e}^{-\beta}=\frac{\alpha-1}{\alpha+1}$ and $N=\left(
\alpha+1\right) /2\alpha$.

We proceed now with the optimization procedure for finding states $\hat{\rho}
$ which satisfy Eqs.(\ref{c1})-(\ref{c2}), (\ref{c3})-(\ref{c4}) and
extremize $\alpha $. For each state $\hat{\rho}$ an eigenbasis exists such
that $\hat{\rho}=\sum c_{n}\left\vert \Psi _{n}\right\rangle \left\langle
\Psi _{n}\right\vert $ \ with $0\leq c_{n}\leq 1$ and ${\sum }c_{n}=1$. We
can also rewrite the state as $\hat{\rho}=\sum \left\vert \psi
_{n}\right\rangle \left\langle \psi _{n}\right\vert $, by using the
unormalized eigenvectors $\left\vert \psi _{n}\right\rangle =\sqrt{c_{n}}\left\vert \Psi _{n}\right\rangle $, while additionally imposing the
normalization constraint 
\begin{equation}
\mathrm{Tr}\left( \hat{\rho}\right) =1.  \label{n1}
\end{equation}
In this way the positivity of $\hat{\rho}$ is ensured since the mixing
coefficients $c_{n}$ are just the squared norms $c_{n}=\left\langle \psi
_{n}\right. \left\vert \psi _{n}\right\rangle $.

The next step is to choose an orthonormal basis $\left\{ \left\vert
i\right\rangle \right\} $ and decompose the vectors $\left\vert \psi
_{n}\right\rangle =\sum \psi _{n}^{i}\left\vert i\right\rangle $. We can
re-express accordingly the uncertainty (\ref{q3}) and constraints (\ref{c1}
)-(\ref{c2}), (\ref{c3})-(\ref{c4}), and (\ref{n1}) as functions of the
complex amplitudes $\psi _{n}^{i}$'s. This gives%
\begin{equation}
\alpha =\sum_{n,i,j}\psi _{n}^{i\ast }\psi _{n}^{j}\left\langle i\right\vert
\left( 2\hat{n}+1\right) \left\vert j\right\rangle  \label{a1}
\end{equation}
and 
\begin{eqnarray}
\sum_{n,i,j}\psi _{n}^{i\ast }\psi _{n}^{j}\left\langle i\right\vert \hat{x}
\left\vert j\right\rangle & =0  \label{c1b} \\
\sum_{n,i,j}\psi _{n}^{i\ast }\psi _{n}^{j}\left\langle i\right\vert \hat{p}
\left\vert j\right\rangle & =0  \label{c1bb} \\
\sum_{n,i,j}\psi _{n}^{i\ast }\psi _{n}^{j}\left\langle i\right\vert \left( 
\hat{x}\hat{p}+\hat{p}\hat{x}\right) \left\vert j\right\rangle & =0
\label{q2b} \\
\sum_{n,i,j}\psi _{n}^{i\ast }\psi _{n}^{j}\left\langle i\right\vert \left( 
\hat{x}^{2}-\hat{p}^{2}\right) \left\vert j\right\rangle & =0  \label{q3b} \\
\sum_{n,i,j}\psi _{n}^{i\ast }\psi _{n}^{j}\left\langle i\right\vert
e^{-\beta \hat{n}}\left\vert j\right\rangle /N& =g \\
\sum_{n,m,i,j}\psi _{n}^{i\ast }\psi _{n}^{j}\psi _{m}^{j\ast }\psi
_{m}^{i}& =\mu \\
\sum_{n,i}\psi _{n}^{i}\psi _{n}^{i\ast }& =1.  \label{n1b}
\end{eqnarray}

The Lagrange multipliers method is well suited as an optimization procedure
for this problem. This method provides necessary conditions on the solution,
which remains invariant under the exchange of any of the constraints with
the quantity to be optimized. Since it is more convenient for us to optimize
over the purity while setting the uncertainty as a constraint, we proceed
accordingly. After differentiating over the amplitudes $\psi _{n}^{i}$ we
obtain the following necessary condition on the eigenvectors $\left\vert
\psi _{n}\right\rangle $

\begin{eqnarray}
\left( \lambda _{0}I+\lambda _{1}\hat{n}+\lambda _{2}e^{-\beta \hat{n}}/N+\lambda _{3}\hat{x}+\lambda _{4}\hat{p}\right. &  \nonumber \\
\left. +\lambda _{5}\left( \hat{x}\hat{p}+\hat{p}\hat{x}\right) +\lambda
_{6}\left( \hat{x}^{2}-\hat{p}^{2}\right) -\hat{\rho}\right) \left\vert \psi
_{n}\right\rangle & =0.  \label{sol}
\end{eqnarray}
The term $\hat{\rho}$ here appears as a consequence of the purity term $\mathrm{Tr}\left( \hat{\rho}^{2}\right) $. This condition can be re-written
as $\left( \hat{H}-\hat{\rho}\right) \left\vert \psi _{n}\right\rangle =0$
where $\hat{H}$ is a Hermitian operator defined as 
\begin{eqnarray}
\hat{H}& =\lambda _{0}I+\lambda _{1}\hat{n}+\lambda _{2}e^{-\beta \hat{n}
}/N+\lambda _{3}\hat{x}+\lambda _{4}\hat{p}  \nonumber \\
& +\lambda _{5}\left( \hat{x}\hat{p}+\hat{p}\hat{x}\right) +\lambda
_{6}\left( \hat{x}^{2}-\hat{p}^{2}\right) .  \label{Ho}
\end{eqnarray}
We can employ the fact that $c_{n}=\left\langle \psi _{n}\right. \left\vert
\psi _{n}\right\rangle $ to express this condition in the form 
\begin{equation}
\hat{H}\left\vert \psi _{n}\right\rangle -c_{n}\left\vert \psi
_{n}\right\rangle =0  \label{Ho2}
\end{equation}
or equivalently as 
\begin{equation}
\hat{H}\left\vert \Psi _{n}\right\rangle -c_{n}\left\vert \Psi
_{n}\right\rangle =0  \label{Ho3}
\end{equation}
One can conclude that the eigenvectors $\left\vert \Psi _{n}\right\rangle $
of the solution $\hat{\rho}$ are the eigenvectors $\left\vert \phi
_{n}\right\rangle $ of the Hermitian operator $\hat{H}$ while the mixing
coefficients $c_{n}$ are the corresponding\textit{\ positive} eigenvalues $\varepsilon _{n}^{+}$ of $\hat{H}$. In other words, the Lagrange multipliers
method provides a necessary condition on the expression of the solution $
\hat{\rho}$. It is written as 
\begin{equation*}
\hat{\rho}=\sum_{n}\varepsilon _{n}^{+}\left\vert \Psi _{n}\right\rangle
\left\langle \Psi _{n}\right\vert .
\end{equation*}

We should note here an important difference between the condition that we
obtain here, (\ref{Ho3}), and the necessary condition obtained in \cite{PRAr}
where all constraints are linear, i.e. can be expressed in the form $\mathrm{Tr}\left( \hat{\rho}B\right) $ with $\hat{B}$ a Hermitian operator. In that
case the condition dictates that the all the eigenvectors of the solution $\hat{\rho}$ should correspond to the same eigenvalue of a Hermitian operator 
$\hat{H}$. The degeneracy constraint is lifted here due to the presence of
the non-linear constraint of the purity $\mathrm{Tr}\left( \hat{\rho}^{2}\right) $. With this more general example, we complete the description
of the method for the derivation of bounded uncertainty relation originally
described in \cite{PRAr}.

One should now proceed with the identification of the eigenvectors of the
Hermitian operator $\hat{H}$, a task that is not that simple because of the
presence of the term $e^{-\beta \hat{n}}/N$ in\ Eq.(\ref{Ho}). The problem
can be simplified, as shown in the Appendix. There it is proven that the
states that the purity possess a phase-independent Wigner function and
therefore \ are confined to be mixtures of number states 
\begin{equation}
\hat{\rho}=\sum_{n}\varepsilon _{n}^{+}\left\vert n\right\rangle
\left\langle n\right\vert .  \label{r}
\end{equation}
Obviously, the solution to the optimization problem consists of states which
either maximize or minimize the purity $\mu $ for fixed uncertainty $\alpha $ \ and Gaussianity degree $g$. As we are going to show at the end, the states
which maximize the purity are not relevant for our purposes here, and thus
we proceed by identifying the states of minimum purity which can be
expressed as in Eq.(\ref{r}).

\begin{figure}[h]
{\centering{\includegraphics*[width=0.45\textwidth]{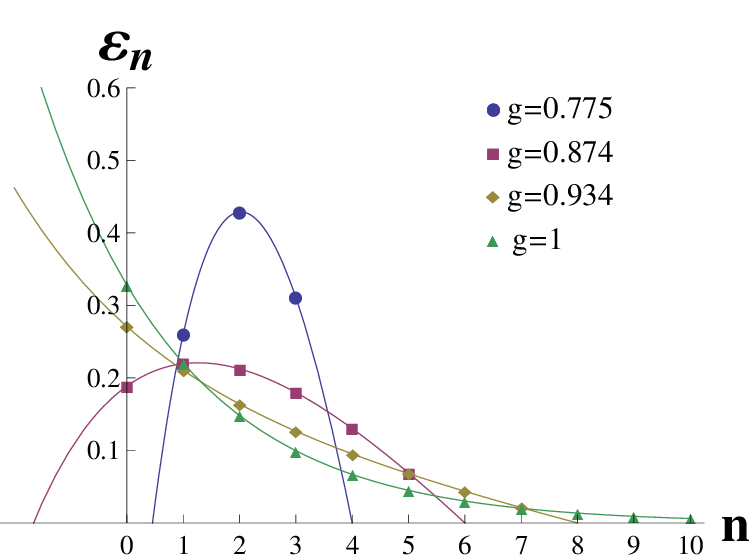}}} 
\vspace{0.1cm}
\caption{Different representative shapes of the positive spectrum $\protect\varepsilon _{n}$ of the Hermitian operator $\hat{H}_{0}$ for $\protect
\alpha =5.1$. The corresponding values of the Gaussianity and purity are $\left( g,\protect\mu \right) =\left\{ \left( 0.775,0.348\right) ,\left(
0.874,0.183\right) ,\left( 0.934,0.176\right) ,\left( 1,0.196\right)
\right\} $}
\label{FigOne}
\end{figure}

Having restricted ourselves to states of the form Eq.(\ref{r}), we ensure
that the constraints Eqs.(\ref{c1b})-(\ref{q3b}) are automatically satisfied
and the restriction of the Hermitian operator $\hat{H}$ on this class of
states becomes 
\begin{equation}
\hat{H}_{0}=\lambda _{0}\hat{I}+\lambda _{1}\hat{n}+\lambda _{2}e^{-\beta 
\hat{n}}/N.  \label{Hob}
\end{equation}
The eigenstates of $\hat{H}_{0}$ are the number states $\left\vert
n\right\rangle $ (for the non-degenerate case) and the corresponding
spectrum is 
\begin{equation}
\varepsilon _{n}=\lambda _{0}+\lambda _{1}n+\lambda _{2}\frac{2\alpha \left(
\alpha -1\right) ^{n}}{\left( \alpha +1\right) ^{n+1}}.  \label{ron}
\end{equation}
where $\lambda $'s are to be identified by the constraints Eqs.(\ref{q3}), \
(\ref{c4}) and (\ref{n1}). The difficult part is to identify among the
eigenstates of $\hat{H}_{0}$ those with positive eigenvalues $\varepsilon
_{n}^{+}$. \ To do so one should first identify the possible structures of
positive spectrum that correspond to different possible values of $\lambda
_{0}$, $\lambda _{1}$ and $\lambda _{2}$ (or equivalently to different
values of $\alpha $ and $g$). In Fig. \ref{FigOne} we present four different
representative `shapes' of the spectrum corresponding to a fixed $\alpha $
and varying $g$. By inserting the positive spectrum $\varepsilon _{n}^{+}$
into Eq.(\ref{r}) one gets straightforwardly the value of the purity $\mu
=\sum_{n}\left( \varepsilon _{n}^{+}\right) ^{2}$.

One can see that for all cases the \textit{positive} spectrum corresponds to 
\textit{\ successive} number states, so we can conclude that the MSs have
the form 
\begin{equation}
\hat{\rho}=\sum_{n_{\min }}^{n_{\max }}\varepsilon _{n}^{+}\left\vert
n\right\rangle \left\langle n\right\vert .  \label{rogen}
\end{equation}
where \ $n_{\min }$ and $n_{\max }$ are parameters which also depend on the
constraints $\alpha $ and $g$ in a complicated fashion.

\section{Bounded uncertainty relation\label{Section3}}

Having identified the form of the MSs we can proceed with the identification
of $\lambda _{0}$, $\lambda _{1}$ and $\lambda _{2}$ in Eq.(\ref{ron}) by
imposing the constraints of normalization, uncertainty $\alpha $ and
Gaussianity $g$. This should be done for all pairs of $n_{\min }$ and $n_{\max }$, but then one should go back and keep only the pairs for which
the eigenvalues $\left\{ \varepsilon _{n_{\min }},\varepsilon _{n_{\min
}},\ldots \varepsilon _{n_{\max }}\right\} $ are positive while all the rest
of eigenvalues are negative. Obviously, for some choices of values of
uncertainty and Gaussianity no pair of $n_{\min }$ and $n_{\max }$
satisfying these conditions, exists. Otherwise, we can in principle deduce
the values of $n_{\min }$ and $n_{\max }$ which are consistent with the
constraints of normalization, uncertainty $\alpha $ and Gaussianity $g$.
This finally yields the extremal purity $\mu $.

In what follows, we expose one possible way for simplifying this complicated
procedure by fixing $n_{\max }$ instead of the Gaussianity $g$. Then we
still have to check all values of $n_{\min }$ and keep those that satisfy
the positive spectrum condition. The key observation is that if Eq.(\ref{ron}) is re-written substituting the discrete index $n$ by a continuous variable 
$x$, 
\begin{equation}
\lambda _{0}+\lambda _{1}x+\lambda _{2}\frac{2\alpha \left( \alpha -1\right)
^{x}}{\left( \alpha +1\right) ^{x+1}}  \label{root}
\end{equation}
then the zeros of this equation (which are maximum $2$ in number) define \ $n_{\min }$ \ and $n_{\max }$. \ More precisely, if the equation has two
positive roots, $x_{1}$ and $x_{2}$ ($>x_{1}$), then $n_{\min }=\left\lceil
x_{1}\right\rceil $, $n_{\max }=\left\lfloor x_{2}\right\rfloor $ (where $\left\lceil x\right\rceil $ is ceiling function and $\left\lfloor
x\right\rfloor $ the floor function). In the case where $x_{1}<0$ or we have
only one root $x_{2}>0$ (see yellow curve in \ref{FigOne}) then $n_{\min }=0$. \ In view of these results, we are able to propose a protocol to obtain in
a systematic way the whole set of MSs \ where the constraint on the
Gaussianity is replaced by a constraint on the second root $x_{2}$ of Eq.(\ref{root}) which indirectly fixes $n_{\max }$ :

\begin{enumerate}
\item Fix \ a value for $\alpha>1$, a non-negative integer value for $n_{\min }$ and a real positive value for $x_{2}$ such that $x_{2}-n_{\min}>1$
.

\item Solve the system of equations 
\begin{align}
\sum_{n=n_{\min }}^{\left\lfloor x_{2}\right\rfloor }\left( \lambda
_{0}+\lambda _{1}n+\lambda _{2}\frac{2\alpha \left( \alpha -1\right) ^{n}}{\left( \alpha +1\right) ^{n+1}}\right) \left( 2n+1\right) & =\alpha
\label{s1} \\
\sum_{n=n_{\min }}^{\left\lfloor x_{2}\right\rfloor }\left( \lambda
_{0}+\lambda _{1}n+\lambda _{2}\frac{2\alpha \left( \alpha -1\right) ^{n}}{\left( \alpha +1\right) ^{n+1}}\right) & =1  \label{s2} \\
\lambda _{0}+\lambda _{1}x_{2}+\lambda _{2}\frac{2\alpha \left( \alpha
-1\right) ^{x_{2}}}{\left( \alpha +1\right) ^{x_{2}+1}}& =0  \label{s3}
\end{align}
for $\lambda _{0},$ $\lambda _{1}$, and $\lambda _{2}$ and obtain $\varepsilon _{n}$ from Eq.(\ref{ron}) as a function of $n_{\min }$, $\alpha $
and $x_{2}$. The Eqs.(\ref{s1})-(\ref{s2}) express the constraints Eq.(\ref%
{q3}) and Eq.(\ref{n1}) respectively while Eq.(\ref{s3}) ensures that $x_{2}$
is the root of Eq.(\ref{root}).

\item Verify using the spectrum provided by Eq.(\ref{ron}) that the lowest
index of positive part of the spectrum is indeed $n_{\min }$. In other words
we check that $\varepsilon _{n_{\min }}\geq 0$ and also that $\varepsilon
_{n_{\min }-1}<0$ if $n_{\min }>0$.
\end{enumerate}

This procedure gives the values of the Gaussianity $g$ and the minimum
purity $\mu $ corresponding to the chosen parameters $\alpha $ and $x_{2}$,
namely 
\begin{align}
g& =\sum_{n_{\min }}^{\left\lfloor x_{2}\right\rfloor }\varepsilon _{n}\frac{2\alpha \left( \alpha -1\right) ^{n}}{\left( \alpha +1\right) ^{n+1}}
\label{para1} \\
\mu & =\sum_{n_{\min }}^{\left\lfloor x_{2}\right\rfloor }\varepsilon
_{n}^{2}.  \label{para2}
\end{align}
This yields one MS. To obtain the whole set of MSs this procedure should be
repeated by varying the parameters $\alpha $ and $x_{2}$.

According to our studies the above procedure always yields $g<1$, meaning
that the derived MSs cannot cover values of $g$ greater than one. For $g=1$
we have the limiting case where $\lambda _{0}=\lambda _{1}=0$, there is no
roots for Eq.(\ref{s3}) \ and the MSs are the so called `thermal' states
(see green curve in Fig.\ref{FigOne}). For $g>1$ there is no combination of $\lambda $'s which gives positive spectrum solution but by extrapolating the
results in \cite{PRAr} one can construct a bound of minimum purity with the
following states 
\begin{equation}
\hat{\rho}=r\hat{\rho}_{G}+\left( 1-r\right) \left\vert n\right\rangle
\left\langle n\right\vert  \label{gg}
\end{equation}
where $n\rightarrow \infty $, $r\rightarrow 1$ while $\mu \left( \hat{\rho}\right) $ is kept constant, and $\hat{\rho}_{G}$ a thermal state
\begin{equation*}
\hat{\rho}_{G}=\sum_{m=0}^{\infty }\frac{2\left( \beta -1\right) ^{m}}{\left( \beta +1\right) ^{m+1}}\left\vert m\right\rangle \left\langle
m\right\vert
\end{equation*}
of purity $\mu \left( \hat{\rho}_{G}\right) =\frac{1}{\beta }
$. The uncertainty $\alpha $ for the MSs in Eq.(\ref{gg}) can be easily
calculated 
\begin{equation}
\alpha =\frac{g}{\left( 2-g\right) \mu },g<1  \label{agm}
\end{equation}
where $g$ \ is the Gaussianity $\ $and $\mu $ the purity of the state $\hat{\rho}$.

\begin{figure}[h]
{\centering{\includegraphics*[width=0.28\textwidth]{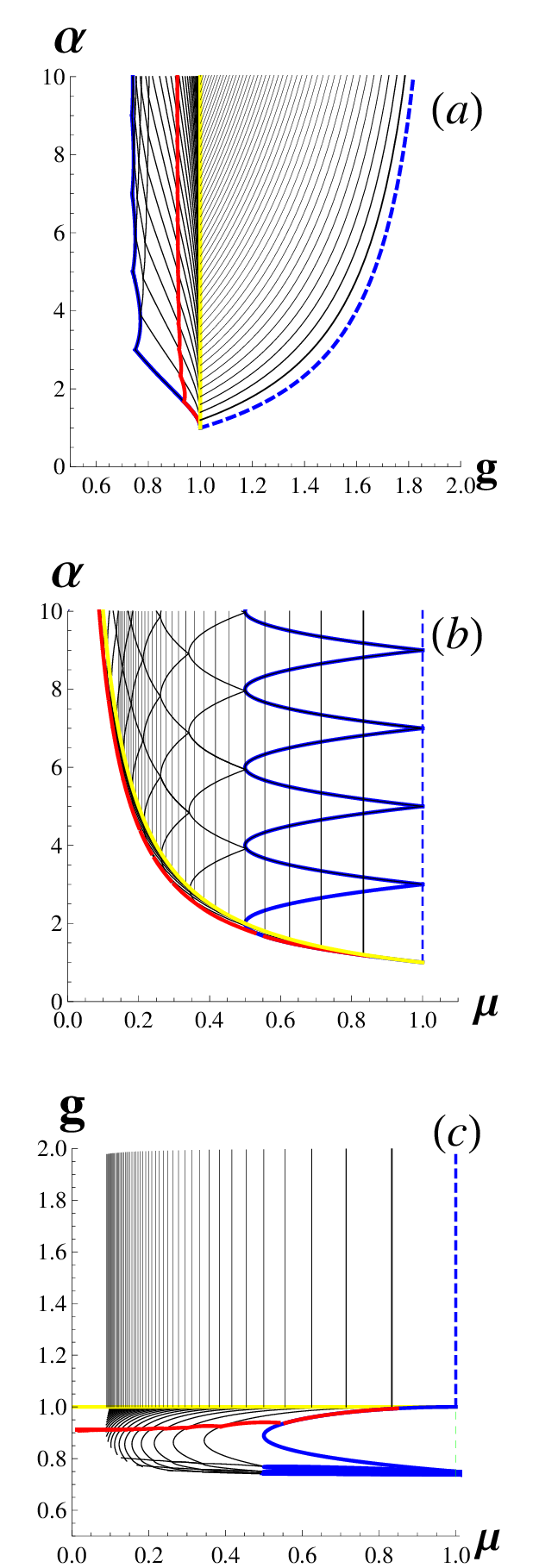}}} 
\vspace{0.1cm}
\caption{The surface that represents the purity and Gaussianity bounded
relation projected on three planes: (a) $\protect\alpha -g$ , (b) $\protect\alpha -\protect\mu $, and (c) $g-\protect\mu $. The red line stands for the
purity-bounded uncertainty relation, the blue (dashed and solid) line for the Gaussianity
-bounded uncertainty relation and the yellow line for the Gaussian states.
For $g>1$ the solid lines are of constant purity (see Eq.(\protect\ref{gg})). For $g<1$ the surface is separated into segments of MSs with fixed $n_{\min}$ 
 and $n_{\max }$ (see Eq.(\protect\ref{rogen})). The lowest and highest value of uncertainty $\alpha$ for each segment define respectively $n_{\min}$ and $n_{\max}$ according to the relation $\alpha=2 n+1$. For the plots we have worked in the domain of uncertainty $1\leq\alpha\leq10$ and for this reason (c) appears incomplete.}
\label{FigTwo}
\end{figure}

\begin{figure}[h]
{\centering{\includegraphics*[width=0.4\textwidth]{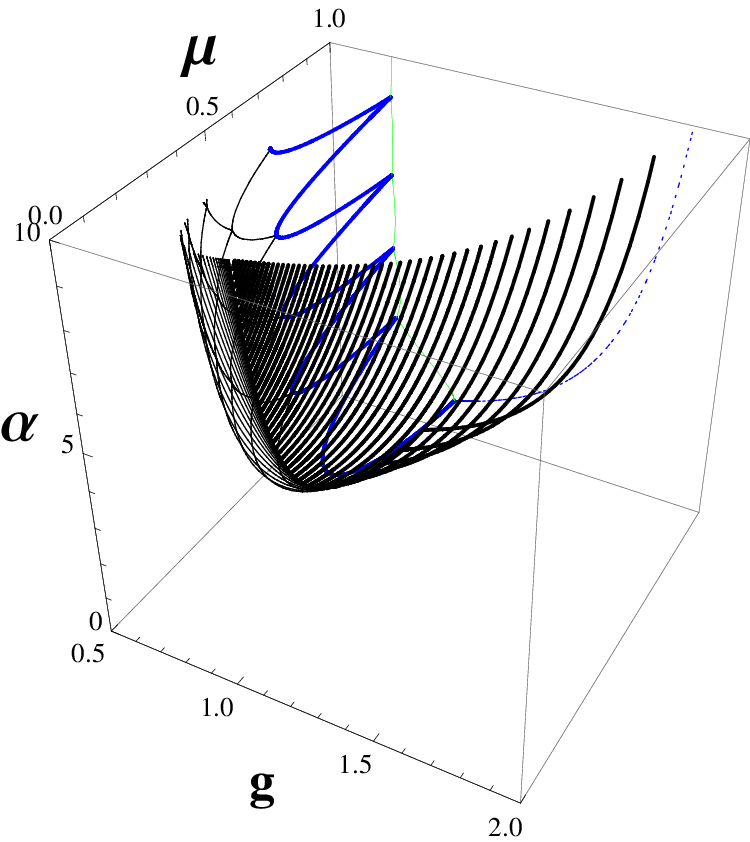}}} 
\vspace{0.1cm}
\caption{The surface that represents the  purity and Gaussianity bounded
relation. For a given value of purity $\protect\mu $ and Gaussianity $g$,
this surface provides the lowest possible value of the uncertainty $\protect
\alpha $ for all quantum states (pure or mixed). The blue line represents the Gaussianity bounded uncertainty relation \cite{PRAr}.}
\label{FigThree}
\end{figure}

One may employ the parametric relations Eqs.(\ref{para1})-(\ref{para2}) \
for $g<1$, to represent graphically the surface that stands for the
Gaussianity and purity bounded uncertainty relation. For $g>1$ one should
employ the much simpler relation given by Eq.(\ref{agm}). \ In Fig.~\ref{FigTwo} we represent the purity and Gaussianity  uncertainty relation
projected on three, mutually orthogonal planes. One can also see on the same
figure the lines which represent the purity bounded uncertainty relation and
the Gaussianity bounded one. These one-dimensional uncertainty relations
appear as outer boundaries (see red and blue line in Fig.~\ref{FigTwo}) of
the surface standing for the purity and Gaussian uncertainty relation. In
Fig. \ref{FigThree} we give a $3-$dimensional view of the uncertainty relation.

One can observe that the surface representing the uncertainty relation is convex for $g>1$
and its boundaries are laying on the plane of pure states.
For $g<1$ a part of the convex surface is somehow ``etched'' by concave grooves
delimited by (blue) loops so that the projections of the loops on $\mu=1$ plane
coincide with  the intersection of the uncertainty surface with the plane.
This reflects the fact that for any given value of $\alpha$ the points of the boundary of the grooves 
have the same value of $g$ for all available values of $\mu$.
Therefore, the curves on the uncertainty surface which correspond to any given $\alpha$ are convex.
This convexity makes sufficient our analysis of the states which minimize the purity
because we perform it independently for all given values of $\alpha$ and
therefore, there is no need to search for the states which maximize the purity. 

Not that, the parametric relation for $g<1$ is not very convenient since it is not
expressed in the desired form $\alpha \geq f\left( g,\mu \right) $ where for
a given value of $g$ and $\mu $ one may conclude on the smallest possible
value on the uncertainty $\alpha $. For this reason we have derived the
following approximate relation for  $f\left( g,\mu \right) $ when $g<1$,
\begin{equation}
f_{\mathrm{app}}\left( g,\mu \right) =\frac{2-g}{g^{7/2}\mu ^{g^{2}}}\left(
1+\frac{0.2\sqrt{\mu }}{\sqrt{0.01\pi }} \mathrm{e}^{ -100(g-0.87)^{2}}
\right) ^{-1}.  \label{app}
\end{equation}
For every pair of values $\left( g,\mu \right) $ it holds that $f\left(
g,\mu \right) >$ $f_{\mathrm{app}}\left( g,\mu \right) $ thus the
approximate formula provides a lower estimation on the real bound of the
uncertainty  $f\left( g,\mu \right) $. \ In view of these results, we
summarize the Gaussianity and purity bounded uncertainty relation as \begin{equation}
\alpha \geq \left\{ 
\begin{array}{cc}
\frac{g}{\left( 2-g\right) \mu } & ,2>g\geq 1 \\ 
f_{\mathrm{app}}\left( g,\mu \right)  & ,1>g\geq 2/e
\end{array}
\right\} .  \label{sum}
\end{equation}

In a previous work \cite{SPIE} we have derived following a similar but less
mathematically consistent method, an uncertainty relation that is bounded by
the degree of von Neumann entropy of a state $\hat{\rho}$  and the quantity
of overlap $\mathrm{Tr}\left( \hat{\rho}\hat{\rho}_{G}\right) $ between the
state and the reference Gaussian state $\hat{\rho}_{G}$. The approach which
we follow here, permits us to assert the positivity of the density matrix of
the solution  in the Lagrange multipliers method while in \cite{SPIE} we
`impose' the positivity on the solution provided by the optimization method.
On the other hand, the   relation obtained in this work in $3-$dimensional
representation,  strongly resembles the one  in \cite{SPIE}, with the main
difference being on the set of MSs. In \cite{SPIE} all MSs are of infinite
rank while here the rank of the solution (the number of eigenvectors of the
solution density matrix) remains finite in the general case ($g\neq 1$).

\section{Conclusions\label{Section4}}

We have introduced and studied an uncertainty relation for the quantum
variables of position and momentum, which is tighter than both the Schr\"{o}dinger-Robertson \cite{Robertson} and the purity bounded uncertainty
relation by Dodonov and Man'ko \cite{Man'ko}. Our new relation makes the
minimum on the  uncertainty \ $\alpha $ a function of the purity $\mu $ of
quantum states and their degree of Gaussianity $g$. Thus the whole \ set of
quantum states of one-dimensional moving particle (or one optical mode)
becomes bounded below in terms of $\alpha $ by a non-trivial surface in
three-dimensional parametric space of $\mu $, $g$, and $\alpha $. Being
projected on the plane $\mu -\alpha $ our bound recovers the purity bounded
uncertainty relation by Dodonov and  Man'ko \cite{Man'ko} while \ its
projection on the $g-\alpha $ plane recovers the Gaussianity bounded
uncertainty relation \cite{PRAr}. Whereas for $g>1$ our surface is given by
an explicit function $\alpha =f\left( g,\mu \right) $ the part of the
surface for $g<1$ is obtained only as a parametric function. In order to
express our result in the desired form for $g<1$ we have constructed an
approximation by function $f_{\mathrm{app}}\left( g,\mu \right) $. This
function determines a surface which for any $(g,\mu )$ lays slightly below
the actual surface: $f\left( g,\mu \right) >f_{\mathrm{app}}\left( g,\mu
\right) $ and thus provides a less tight, but still valid, bound. Finally,
our results allow us to visualize the whole set of quantum states in three
dimensional parametric space and accurately bound the uncertainty of $x$ and 
$p$ taking into account the purity and Gaussianity the states for which the
uncertainty is evaluated.

\bigskip

\acknowledgments AM gratefully acknowledges financial support from the
Belgian National Fund for Scientific Research (FNRS). This work was carried
out with the financial support of the European Commission via projects
HIPERCOM, the support of the Belgian Federal program IAP via the P7/35
Photonics@be project, and the support of the Brussels-Capital Region, via
project CRYPTASC.

\appendix

\section{Appendix}

\textit{Proposition: The states of minimum purity\ for given uncertainty and
Gaussianity can be expressed as mixtures of number states}

Let us consider a general density matrix $\hat{\rho}$ of purity $\mu $ and
Gaussianity $g$, whose covariance matrix has been set via LCT proportional
to the unity \ (in Williamson form) and its displacement vector $\mathbf{d}$
to zero. \ In this case the uncertainty $\alpha $ of the state completely
characterizes the reference Gaussian state $\hat{\rho}_{G}$ which is just a
thermal state. In the Wigner representation $\ $the reference state, $\hat{\rho}_{G}$, 
\begin{equation}
W_{G}\left( r\right) =\frac{1}{\pi \alpha }e^{-r^{2}/\alpha },\qquad r=\sqrt{x^{2}+p^{2}}  \label{Wg}
\end{equation}
has no dependence on the angular degree of freedom $\varphi $. In contrast,
in the general case the state $\hat{\rho}$ itself possess an
angular-dependent Wigner function $W\left( r,\varphi \right) $ and its
purity can be expressed via $W\left( r,\varphi \right) $ as 
\begin{equation}
\mu =2\pi {\displaystyle\iint }W\left( r,\varphi \right) ^{2}r\mathrm{d}r \mathrm{d}\varphi
\end{equation}
while its Gaussianity as 
\begin{equation}
g=2\pi \alpha {\displaystyle\iint }W\left( r,\varphi \right) W_{G}\left(
r\right) r\mathrm{d}r\mathrm{d}\varphi .  \label{gu}
\end{equation}
The next step is to prove that for any given state $\hat{\rho}$, another
state $\hat{\rho}_{s}$ exists of the same $\alpha $ and $g$ and smaller or
equal purity, which possess a phase-independent Wigner function. Let us
define this new state $\hat{\rho}_{s}$ by phase-averaging the Wigner
function $W\left( r,\varphi \right) $ of the initial state $\hat{\rho}$ \ 
\begin{equation}
W_{s}\left( r\right) =\frac{1}{2\pi }\int W\left( r,\varphi \right) d\varphi
.  \label{Ws}
\end{equation}
Here $W_{s}\left( r\right) $ is the Wigner function of the new state $\hat{\rho}_{s}$. The reference Gaussian state of $\hat{\rho}_{s}$ (and
consequently the uncertainty $\alpha $) is the same \ as for $\hat{\rho}$,
since phase-averaging cannot affect the angular-independent Wigner function
Eq.( \ref{Wg}). The Gaussianity degree Eq.(\ref{gu}) remains the same, as
well. This is a straightforward result of substitution of the
phase-independent Wigner function given by Eq.~(\ref{Ws}) into Eq.~(\ref{gu}). On the other hand, the purity $\mu _{s}$ of the symmetrized state $\hat{\rho}_{s}$ is constrained to be smaller than, or equal to, that of $\hat{\rho}$. Indeed, by applying the Cauchy-Schwarz inequality we have 
\begin{align*}
\mu _{s}& =2\pi {\displaystyle\iint }W_{s}\left( r\right) ^{2}r\mathrm{d}r
\mathrm{d}\varphi \\
& ={\displaystyle\iiint }W\left( r,\varphi \right) W\left( r,\Phi \right) r
\mathrm{d}rd\varphi d\Phi \\
& \leq \sqrt{{\displaystyle\iiint }W\left( r,\varphi \right) ^{2}r\mathrm{d}
rd\varphi d\Phi }\sqrt{{\displaystyle\iiint }W\left( r,\Phi \right) ^{2}r
\mathrm{d}rd\varphi d\Phi } \\
& =2\pi {\displaystyle\iint }W\left( r,\varphi \right) ^{2}r\mathrm{d}r
\mathrm{d}\varphi =\mu .
\end{align*}
This concludes the proof of this proposition.

From this proposition it is straightforward to deduce that the minimizing
states we are looking for, are states of angular-independent Wigner function
and therefore can be expressed \cite{Werner} as a convex combination of the 
\textit{number} (Fock) states 
\begin{equation}
\hat{\rho}=\sum \rho _{n}\left\vert n\right\rangle \left\langle n\right\vert
.  \label{rho}
\end{equation}

\end{document}